\documentclass[conference]{IEEEtran}
\usepackage[left=0.625in, right=0.625in, bottom= 1.1in, top=1.0in]{geometry}
\usepackage{graphicx}
\usepackage{balance}
\usepackage{nopageno}
\usepackage{cite}
\usepackage{amsthm}
\usepackage{amssymb}
\usepackage{amsmath}
\usepackage{amsfonts}
\usepackage{amsmath}
\usepackage{graphicx}
\usepackage{mathtools}
\usepackage{multirow}
\usepackage{epstopdf}
\usepackage{subfigure}
\usepackage{xcolor}
\usepackage{array}
\usepackage{csquotes}
\graphicspath{{Figures/}}
\usepackage{mathrsfs} 
\usepackage{listings}
\usepackage{amsmath,relsize,mathtools}

\usepackage[linesnumbered,ruled,vlined]{algorithm2e}

\begin{document}


\title{Evaluation of the effects of  3GPP-specific beamforming and channel estimation on the 3D EIRP profile of a 5G gNB}

\author{ 
\IEEEauthorblockN{
Armed Tusha\IEEEauthorrefmark{1},
Joshua Roy Palathinkal\IEEEauthorrefmark{1}, and
Monisha Ghosh\IEEEauthorrefmark{1}\\
\IEEEauthorblockA{
\IEEEauthorrefmark{1}Department of Electrical and Electronics Engineering, University of Notre Dame, IN, USA.\\
Email: atusha@nd.edu, jpalathi@nd.edu, mghosh@nd.edu}}\\\vspace*{-0.80cm}}
 
\maketitle

\begin{abstract}
Spatial domain exploitation through 3D beamforming serves as a critical technology enabler for performance enhancement in the Fifth Generation New Radio (5G~NR) specification.
This is realized at the gNodeB (gNB) through the integration of massive antenna element arrays that facilitates 3D spatial multiplexing. 
However, these systems with high-directional transmissions also represent a threat to incumbent services such as radar and satellites. These incumbents already operate in mid-band spectrum---including the 4.4--4.9 GHz and 7.125--7.4 GHz bands---that are currently being evaluated for future cellular deployments.
Here, we present the first work that
evaluates the transmitted Effective Isotropic Radiated Power (EIRP) of a gNB in 3D space, using the 3GPP Release-18 standard for FR-1 instead of theoretical analyses of beam nulling, which can be simplistic. We shed light on the problems requiring attention with the EIRP profile in 3D space for existing codebook designs predefined in 3GPP: i) interference from a gNB does not depend only on the worst-case beamforming direction, but on a variety of beamforming directions due to side-lobes; ii) advanced antenna systems (AAS) architecture and antenna port configurations play a crucial role in average 3D EIRP, which are implementation dependent, and iii) we introduce two beam nulling methods, which achieve a 11 dB power reduction toward a target direction, with 3.5--4.5 dB SNR loss in UE link performance at a $10^{-4}$ bit error rate (BER) across modulation schemes under ideal and practical channel estimation, a higher loss compared to predictions from theoretical analyses. 
\end{abstract}


\IEEEpeerreviewmaketitle

\section{Introduction} \label{introduction}

Mid-band spectrum (1--7~GHz) offers a balance between wide coverage and high capacity due to its favorable propagation characteristics and thus plays a key role in enabling both 5G New Radio (5G~NR) communications and radar operations~\cite{5gamericas_midband_spectrum_2021}. The 3~GHz band (bands n77 and n78) is already standardized by 3GPP and commercially mature~\cite{3gpp_38101_1}. The 7~GHz band (7.125--7.4~GHz) is under study for future 5G-Advanced/6G and coexistence with radar and satellite systems~\cite{10459211}.

In 5G netwroks, the transition to higher frequency bands (FR-1 mid-band and FR-2 mmWave), coupled with increased path loss and the need for higher capacity, has made massive Multiple-Input Multiple-Output (mMIMO) a key enabler of 5G performance \cite{bogale2016massive}. In 5G base stations---i.e., the gNodeB (gNB)---the antenna panels are designed to form highly directional beams toward specific user equipment (UE).
This spatially focused transmission enhances the Effective Isotropic Radiated Power (EIRP) in the intended direction, reducing inter-user interference and allowing flexible control of the beam shape in both the azimuth and elevation planes. 


In shared spectrum bands, a key challenge is ensuring coexistence between commercial networks and incumbent systems, particularly cellular networks and radar systems~\cite{ross2019annual}. On September 30, 2025, representatives from the wireless and aviation sectors jointly presented a technical consensus to the Federal Communications Commission (FCC) outlining coexistence parameters for upper C-band (3.98–4.2 GHz) deployments and protection of radio altimeters operating in the adjacent 4.2--4.4~GHz band~\cite{fcc_upper_cband_consensus_2025}. Moreover, as cellular services aim to expand into the upper mid-band (7--24~GHz), they must also coexist with the rapidly proliferating commercial satellite systems~\cite{9681631}.  
In~\cite{farshchian2023modeling}, authors show that the impact of cellular uplink aggregate interference on radar performance follows a non-Gaussian distribution, underestimating its impact on radar systems if treated as additive white Gaussian noise (AWGN). Kang \textit{et al.} investigate cellular-to-satellite interference in the upper mid-band and introduce a beamforming approach by leveraging ephemeris data for null-steering toward visible satellites, protecting the non-terrestrial networks (NTN) link under a negligible SNR penalty for terrestrial downlink~\cite{kang2024terrestrial}. Jo \textit{et al.} present a precoding codebook performing adaptive nulling towards fixed wireless service (FWS) via space-division multiple access for spectrum sharing~\cite{jo2011transmit}. In~\cite{jia2025joint}, authors propose an approach in which a terrestrial base station uses the preamble signals broadcast by non-terrestrial UEs and performs beamforming with spatial nulls toward the victims while serving the desired UE.

A commercial gNB employs large antenna arrays, commonly known as Advanced Antenna Systems (AAS), leveraging UE-specific beamforming to enhance  the link performance~\cite{ericsson_aas_5g_2023, idst_aas_breakthroughs_2022}. gNB vendors often customize the AAS design, including the number of antenna elements, subarrays, and panel size, for specific deployment scenarios such as urban, suburban, or rural environments. These site-specific configurations produce diverse three-dimensional (3D) EIRP patterns in both azimuth and elevation. Such highly directional transmissions can inadvertently generate interference to incumbent systems, since 5G~NR beamforming techniques are primarily optimized for UE performance rather than incumbent protection. 

In 5G~NR, first, the gNB transmits a set of Synchronization Signal Blocks (SSBs) over time and/or frequency using predefined beam sweeping for cell search and initial access. Later, the gNB selects a precoding matrix (PM) from predefined codebooks based on the channel state information (CSI) feedback of the UE, which includes the precoding matrix indicator (PMI), rank indicator (RI), and channel quality indicator (CQI) over the region seen by the selected SSB. However, the gNB is not required to use the reported PMI and selects an optimal PM to form narrow, user-specific beams that maximize signal quality while minimizing interference among connected UEs. The prior art is mainly theoretical and does not account for these practical constraints when following 3GPP standards.

Interference from a gNB to incumbent systems is strongly dependent on the characteristics of the physical AAS architecture, channel estimation, and the selected PM at gNB. To the best of the authors’ knowledge, the 3D EIRP of a commercial gNB and its implications for coexistence with incumbents have not yet been investigated in the literature. The main contributions of this study are summarized as follows:

\begin{itemize}

\item  We analyze the normalized EIRP emitted by a practical gNB in 3D space as a function of the AAS panel design and the standard 3GPP Type I single-panel codebook configuration used for UE-specific beamforming in 5G NR.

\item We evaluate the trade-offs between the average EIRP pointing at a fixed direction in 3D space and the UE's link reliability with ideal and practical CSI for coexistence with airborne incumbents in shared spectrum.

\item We introduce two 3GPP-compliant beam-nulling methods that reduce radiation toward the target direction by 11~dB, while causing  a 3.5--4.5~dB SNR loss at a UE BER of $10^{-4}$ across various modulation schemes.



\color{red}



\end{itemize}
\color{black}
\section{Active Antenna System in 5G }\label{AASsysmodel} 
The 3GPP TR38.901  has numerous options for beamforming, many of which are implementation-dependent~\cite{3gpp_tr_38901_2022}. Manufacturers can select site-specific options, which are usually proprietary, both in terms of gNB physical architecture and codebook-based beamforming selection.
\begin{figure}
	\centering
        \begin{subfigure}[Normalized radiation power pattern of antenna element.]
	    {\includegraphics[scale=.3]{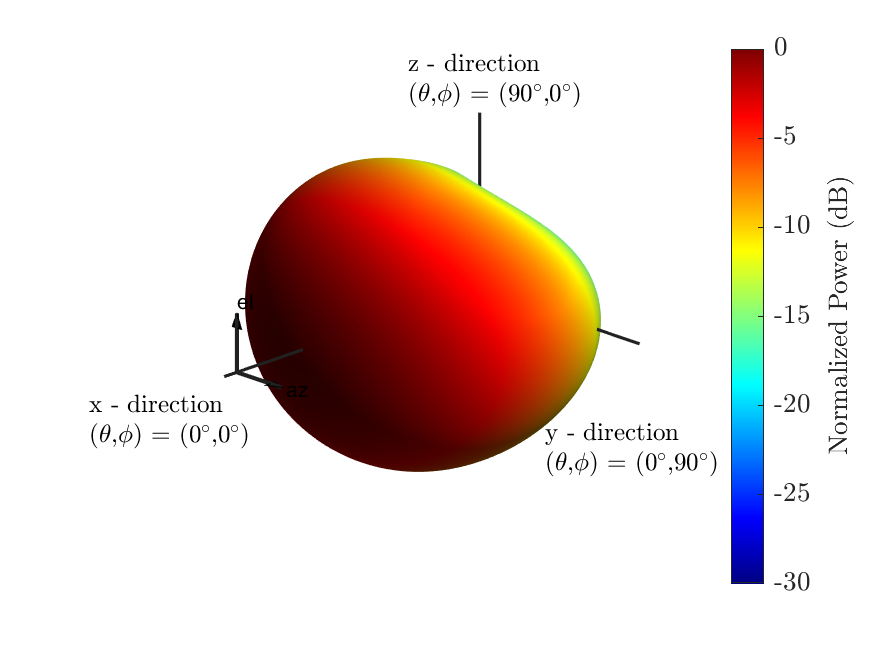}
	    	\label{Fig:antEle-3D}}
	\end{subfigure}
    \begin{subfigure}[Subarray dimentions af AAS, ($M_1$,$M_2$) = (2,3).]
	    {\includegraphics[scale=.2]{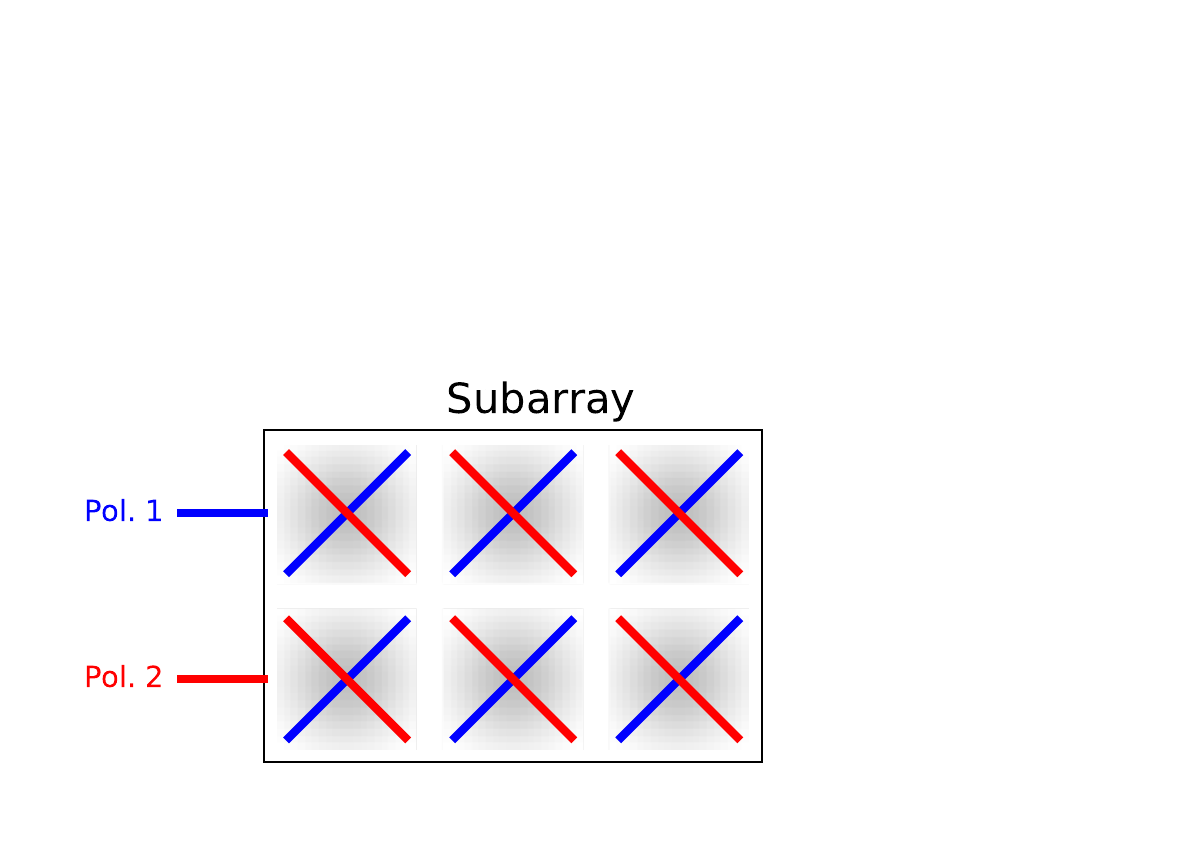}
	    	\label{Fig:subarrayDim}}
	\end{subfigure}
    \begin{subfigure}[Antenna array dimenstions of AAS, ($N_1$,$N_2$) = (4,4)]
	{\includegraphics[scale=0.36]{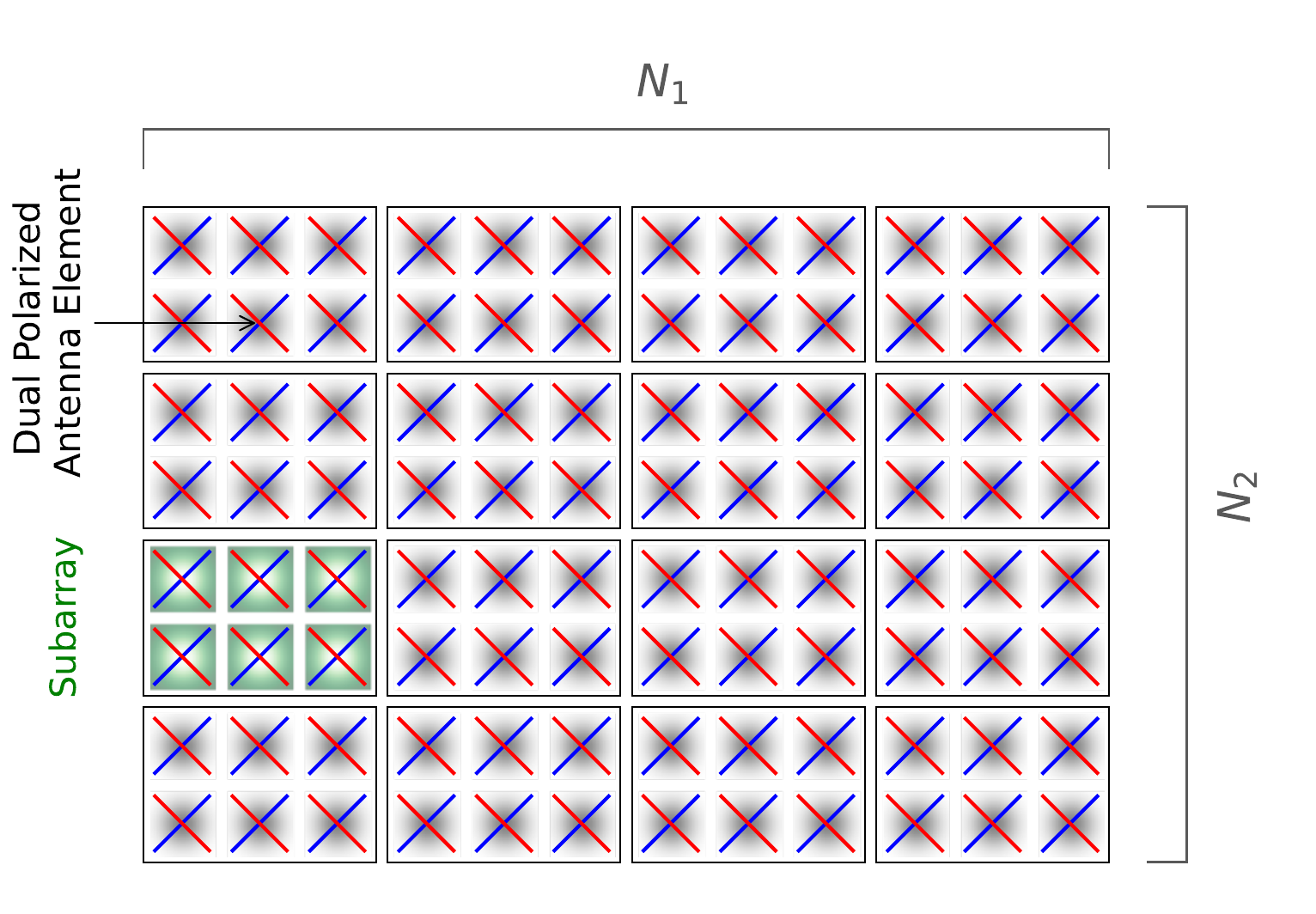}\label{Fig:arrayDim}}
	\end{subfigure}
    \caption{Physical architecture of AAS panel.}

	\label{Fig:ASSarchitecture}
\end{figure}

\subsection{AAS Structure and Parameters}


The International Telecommunication Union (ITU) AAS array model is used to model the gNB's AAS as illustrated in Fig. \ref{Fig:ASSarchitecture}. It consists of three main components: (1) dual-polarized antenna elements, (2) subarrays, and (3) antenna array.

\textit{1) Antenna Element} is the basis unit of the AAS array and is characterized by its operating frequency, polarization, half-power beamwidth (HPBW), and directional gain. The antenna polarization is denoted by $P$, with $P=1$ for single and $P=2$ for cross-polarization. 
In this work, we used the NR antenna element in MATLAB, a dual-polarized directional antenna element capable of  transmitting and receiving of two orthogonally polarized signals at \(\pm45^{\circ}\)~\cite{3gpp_tr_38901_2022}. The normalized radiation power pattern of the NR antenna element is shown in Fig. \ref{Fig:antEle-3D}, and the corresponding configuration parameters are listed in Table~\ref{tab:antElement}.


\begin{table}[!t]
\caption{Key parameters of the NR antenna element}
\centering
\begin{tabular}{|p{0.42\columnwidth}|p{0.48\columnwidth}|}
\hline
\multicolumn{2}{|c|}{\textbf{Antenna Element Characteristics}} \\ \hline
Frequency range & 3700--3980~MHz \\ \hline
Carrier frequency & 3.75~GHz \\ \hline
HPBW in azimuth plane & $90^\circ$ \\ \hline
HPBW in elevation plane & $60^\circ$ \\ \hline
Polarization & Dual-polarized \\ \hline
Directional gain & 5.3~dBi \\ \hline
\end{tabular}
\label{tab:antElement}
\end{table}

\textit{2) Sub-array} is a group of adjacent antenna elements driven by a single RF chain. In our implemented AAS array, each subarray consists of six dual-polarized antenna elements arranged as three pairs of two antenna elements in two rows. Number of rows and number of columns for the subarray are denoted by \(M_{1}\) and \(M_{2}\), respectively, as illustrated in Fig.~\ref{Fig:subarrayDim}. In this work, each two adjacent antenna element pairs are spaced by \(d_{el_v}=0.058\)~m along the vertical plane and \(d_{el_h}=0.044\)~m along the horizontal plane.

\begin{figure}
	\centering
\includegraphics[scale=0.5]{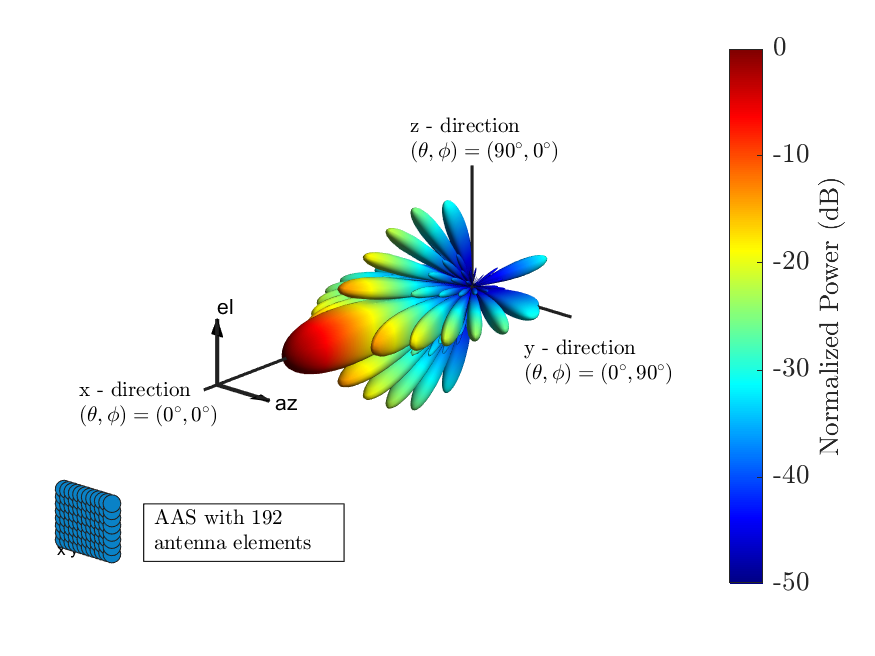}
	\caption{Normalized radiation power pattern (relative EIRP) from AAS for PMI-based precoding.}
	\label{Fig:antEle}
\end{figure}

\textit{3) Antenna Panel Array} is created by grouping multiple subarrays and can be implemented as a single panel or as a multi-panel configuration~\cite{asplund2020advanced}. In this work, the ITU AAS array is constructed by arranging subarrays, each consisting of twelve antenna elements, in a 4\(\times\)4 grid across the horizontal (\(N_{1}\)) and vertical (\(N_{2}\)) planes, resulting in a total of 192 polarized antenna elements or 96 dual-polarized antenna elements, as shown in Fig.~\ref{Fig:arrayDim}. These subarrays are spaced apart by $d_{su_v}=0.174$~m vertically and $d_{su_h}=0.044$~m horizontally. Antenna panel can be configured as either a uniform linear array (ULA) or a uniform planar array (UPA). We use a single UPA panel to model the gNB's AAS. The normalized radiation power pattern of the antenna panel array is shown in Fig.~\ref{Fig:antEle}, with half power beam width (HPBW) \(\theta_{HPBW}=8.74^{\circ}\)  and \(\phi_{HPBW}=7.68^{\circ}\) along elevation (\(\theta\)) and azimuth (\(\phi\)) axis, respectively; its corresponding parameters are given in Table~\ref{tab:panelarray}.

\subsection{Port Mapping According to 3GPP Codebook}

For UE-specific beamforming in 5G~NR, Channel State Information Reference Signals (CSI-RS) can be transmitted through up to 32(FR-1)/64(FR-2) antenna ports, each representing a distinct channel to be sounded. An antenna port is defined such that the channel experienced by one symbol can be inferred from another symbol on the same port, enabling accurate channel estimation. 

In 3GPP TS 38.214, the mapping between transmission layers and antenna ports is defined by the PM $\mathbf{W} \in \mathbb{C}^{P_{\text{CSI-RS}} \times N_L}$, where $P_{\text{CSI-RS}}$ and $N_L$ represent the number of CSI-RS antenna ports and the number of transmission layers, respectively~\cite{gNBH2}. Four types of codebooks are defined to support different antenna configurations and beamforming requirements. Type~I single-panel and multi-panel codebooks are typically used in FR1, while Type~II and enhanced Type~II codebooks are primarily designed for massive MIMO deployments in FR2. 

\begin{table}[!t]
\caption{Key parameters of the AAS array} 
\centering
\begin{tabular}{|p{0.48\columnwidth}|p{0.42\columnwidth}|}
\hline
\multicolumn{2}{|c|}{\textbf{Antenna Array Parameters}} \\ \hline
Antenna Element & NR antenna element~\cite{3gpp_tr_38901_2022} \\ \hline
Vertical/horizontal antenna element spacing within subarray & 0.058/0.044 m \\ \hline
Vertical/horizontal subarray spacing  & 0.174/0.044 m \\ \hline
Sub-array dim. (\(M_{1},M_{2}\)) & 2 $\times$ 3\\ \hline
Panel dim. (\(N_{1},N_{2}\)) & 4 $\times$ 4 sub-arrays \\ \hline
Panel direction & x-axis \\ \hline
Panel down-tilt & $0^\circ$ \\ \hline
\end{tabular}
\label{tab:panelarray}
\end{table}

In this work, the MATLAB 5G Toolbox is used to configure beamforming options for FR1. The AAS array, consisting of 192 antenna elements, is mapped to 32 antenna ports for the Type~I single-panel codebook. ($N_1$,$N_2$) = (4,4) results in a total of $P_{CSI-RS} = N_1 \times N_2 \times 2 = 32$ antenna ports. Thus, each antenna port corresponds to one polarization of subarray within the panel in our simulation. Note that $(N_1, N_2)$ can also be configured as (8,2) or (16,1) to achieve $P_{\text{CSI-RS}} = 32$ ports, resulting in different azimuth and elevation resolutions. In this work, PM weights are applied to the antenna ports column-wise, top to bottom, across all four columns of the AAS. This mapping is implementation dependent and, accordingly, changes. 

\begin{figure}
	\centering
\includegraphics[scale=0.5]{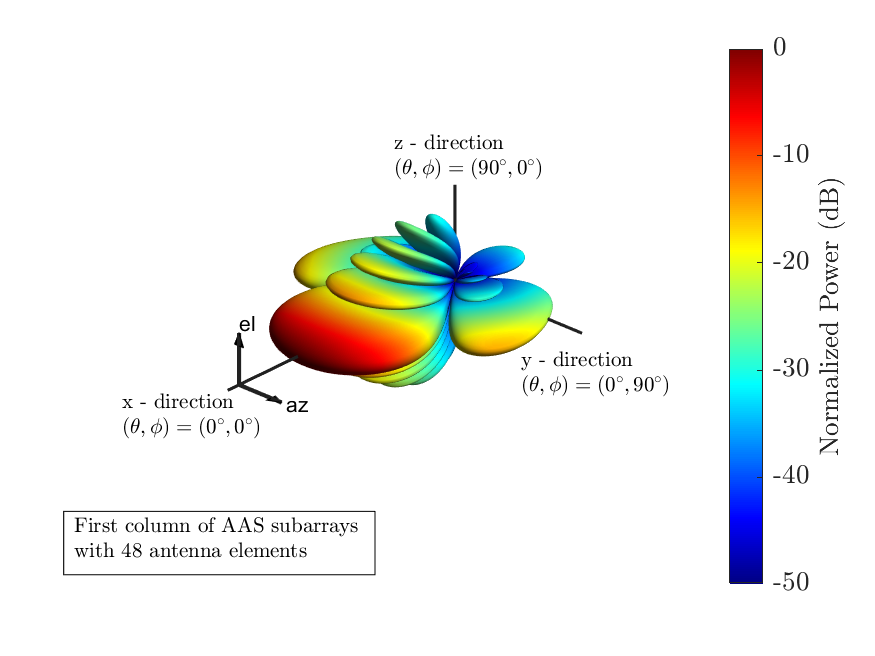}
	\caption{Normalized radiation power pattern of a given SSB.}
	\label{Fig:ssbPatern}
\end{figure}
\subsection{Beamforming for Initial Access}
In contrast to UE-specific PM-based beamforming, the SSB-based beamforming used to perform cell search and initial access beams is not standardized or codebook-based; instead, its implementation is vendor-specific and transparent to the UE's location. Consequently, while PM-based beamforming relies on CSI-RS sounding and UE feedback to select standardized precoding matrices for data transmission, SSB beamforming is a network-driven sweeping mechanism defined by transmission structure and indexing rather than explicit precoding matrix signaling. Fig.~\ref{Fig:ssbPatern} shows the SSB that is generated using all the first column subarrays of AAS. The HPBW characteristics of the SSB are \(\hat{\theta}_{HPBW}=8.74^{\circ}\)  and \(\hat{\phi}_{HPBW}=30.7^{\circ}\), where placing 4 SSBs along the \(\phi\)-plane can cover the sector of a regular three-sector-sized cellular cell. It is worth noting that the average relative EIRP of AAS depends not only on the PM set but also on the spatial characteristics of the SSB gain pattern. In this work we consider 8 SSB-s with steering angle (\(\theta_{SSB,i},\phi_{SSB,i}\)) at three different elevation angles as \(\theta_{SSB,i}\in[6^\circ, 0^\circ, -3^\circ]\), with \(\phi_{SSB,i}\in[-60^\circ, 0.5^\circ, 60.5^\circ]\) for \(\theta_{SSB,i}=6^\circ\) and \(\theta_{SSB,i}=0^\circ\), while \(\phi_{SSB,i}\in[-45^\circ, 45^\circ]\) for \(\theta_{SSB,i}=-3^\circ\).






\section{Proposed 3GPP-Compliant Beam Nulling Schemes }
This section first presents the downlink transmission model in 5G~NR, highlighting the effect of PM, and then introduces two different 3GPP-compliant beam nulling schemes.

\subsection{Downlink Transmission Model}\label{subsec:system_model}
A gNB with $N_T$ transmit antennas communicates with a UE equipped with $N_R$ receive antennas. In this work, we assume $N_T=P_{CSI-RS}$ for the link-level simulator. The received signal at the $i$-th antenna of the UE  is the sum of the contributions from all CSI-RS antenna ports of the gNB
\begin{equation}
    y_i = \sum_{p=0}^{P_{\text{CSI-RS}}-1} h_{i,p} \left( \sum_{l=0}^{N_L-1} W_{p,l} s_l \right) + n_i, \quad i = 0, \dots, N_r-1,
    \label{Eq:1}
\end{equation}
where $s_l$ represents the transmitted symbol on the $l$-th layer with $l\in \{0,\cdots,N_L-1\}$. $W_{p,l}$ is the precoding coefficient mapping the $l$-th layer onto the $p$-th antenna port, with $p\in \{0,\cdots,P_{CSI-RS}-1\}$. $h_{i,p}$ denotes the channel fading coefficient between the $p$-th port and the $i$-th receiver antenna while $n_i$ is the AWGN with $ \mathcal{CN}(0, \sigma^2  )$. Each layer is thus transmitted as a linear combination of multiple antenna ports. The matrix representation of (\ref{Eq:1}) is given as follows
\begin{equation}
    \mathbf{y} = \mathbf{H}_{\text{CSI-RS}} \mathbf{W} \mathbf{s} + \mathbf{n},
\label{Eq:2}
\end{equation}
where $\mathbf{s} \in \mathbb{C}^{N_L \times 1}$ represents the transmitted data symbols for each layer over the physical downlink shared channel (PDSCH), and $\mathbf{W} \in \mathbb{C}^{P_{\text{CSI-RS}} \times N_L}$ is the PM used at the gNB. $\mathbf{y} \in \mathbb{C}^{N_r \times 1}$ denotes the received signal vector after the addition of AWGN, modeled as $\mathbf{n} \in \mathbb{C}^{N_r \times 1} \sim \mathcal{CN}(0, \sigma^2 \mathbf{I})$.

After performing orthogonal frequency division multiplexing (OFDM), the precoded signal propagates through the channel $\mathbf{H}_{\text{CSI-RS}} \in \mathbb{C}^{N_r \times P_{\text{CSI-RS}}}$ between the gNB and UE. The gNB transmits CSI-RS and the UE performs the channel estimation, while reporting back PMI.  The statistical characteristics of the channel responses depend on the selected channel model. In this work, we modeled $\mathbf{H}_{\text{CSI-RS}}$ as a tapped delay line (TDL-C) MIMO fading channel, corresponding to the Urban Macro NLOS scenario\cite{3gpp_tr_38901_2022}. Moreover, the PM corresponding to the reported PMI is used as a PDSCH precoder. 
 
At the receiver, timing synchronization is performed using the demodulation reference signal (DMRS), followed by OFDM demodulation and minimum mean square error (MMSE) equalization. The soft-symbol log-likelihood ratios are then processed through the DL-SCH decoder with normalized min-sum LDPC decoding to recover the transmitted bits.

\subsection{3GPP-Compliant Beam Nulling}
In this work, to control the average EIRP in the direction of interest ($\theta_{i},\phi_{i}$) we propose two different 3GPP-compliant beam nulling schemes given as Algorithm~\ref{alg:pm_selection} and Algorithm~\ref{alg:3dB_selection}. 

\begin{algorithm}[t]
\caption{Threshold-based PM Selection}
\label{alg:pm_selection}
\KwIn{$\mathcal{C}_{\text{panel}}$, $\mathcal{C}_{\text{cb}}$, $\varepsilon$, $\theta_{i}$, $\phi_{i}$}
\KwOut{$result$}

\SetKwFunction{FMain}{SubsetPM}
\SetKwProg{Fn}{Function}{:}{}
\Fn{\FMain{$\mathcal{C}_{\text{panel}}$, $\mathcal{C}_{\text{cb}}$,  $\varepsilon$, $\theta_{i}$,$\phi_{i}$}}{
    $W \leftarrow {\small \texttt{getPMIType1SinglePanelCodebook} (\mathcal{C}_{\text{cb}})}$\; 
    $EIRP \leftarrow \texttt{pattern}(\mathcal{C}_{\text{panel}}, W, \theta_{i}, \phi_{i})$\; 
    \eIf{$EIRP_{\theta_{i},\phi_{i}} < \varepsilon$}{
        $result \leftarrow W$\;
    }{
        \texttt{continue}\;
    }
    
    \Return{$result$}\;
}
\end{algorithm}
\subsubsection{Threshold-based PM Selection}
In Algorithm~\ref{alg:pm_selection}, PM selection ensures that the EIRP in the target direction ($\theta_{i},\phi_{i}$) does not exceed the threshold  ($\varepsilon$). $\mathcal{C}_{\text{panel}}$ denotes the antenna panel configuration (Tables \ref{tab:antElement} and Table \ref{tab:panelarray}). 
$\mathcal{C}_{\text{cb}}$ represents the codebook configuration. 
Line~2 retrieves all Type~I PM-s for a single panel; lines~3 computes the EIRP at (\(\theta_{i},\phi_{i}\)); and lines~4-8 discard PM-s whose EIRP exceeds $\varepsilon$. 

\begin{figure*}
	\centering
    \begin{subfigure}[PM-s, ($N_1$,$N_2$) = (2,2)]
	    {\includegraphics[scale=.4]{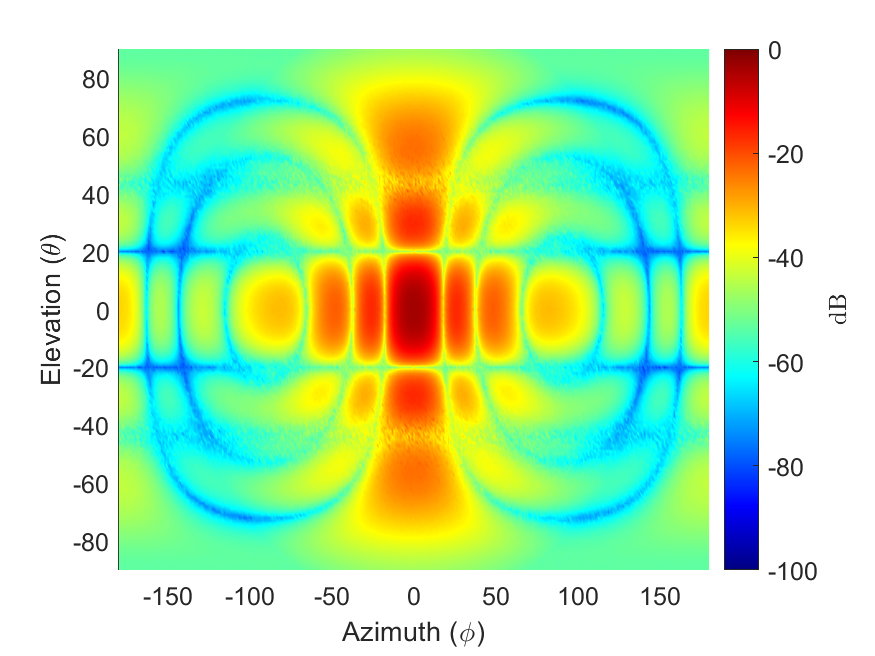}
	    	\label{Fig:EIRP-N1-8-N2-2}}
	\end{subfigure}
    \begin{subfigure}[PM-s, ($N_1$,$N_2$) = (4,4)]
	{\includegraphics[scale=0.4]{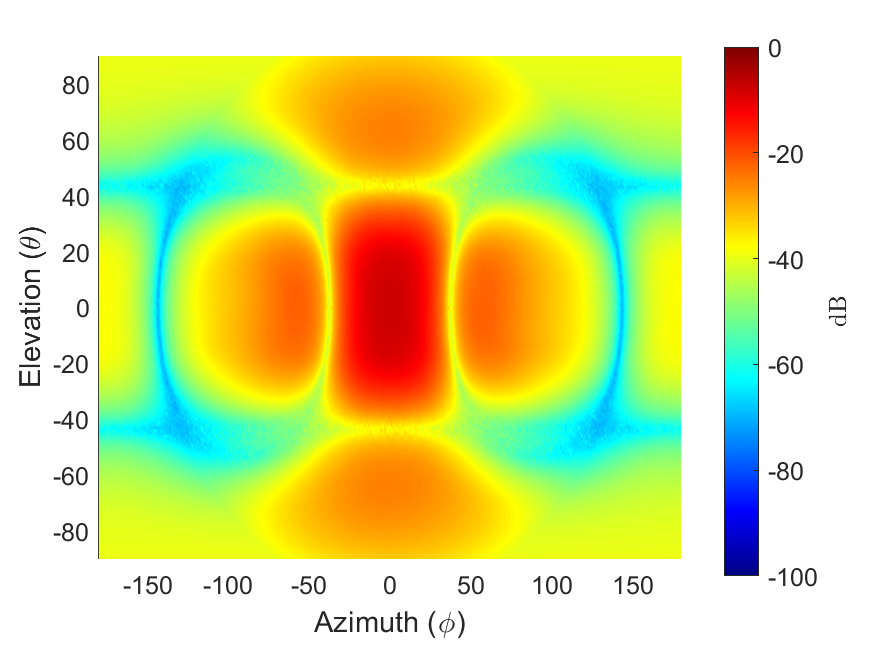}\label{Fig:EIRP-N1-4-N2-4}}
	\end{subfigure}
        \begin{subfigure}[SSB-PM: SSB with conf. \{3, 3, 2\} \& all PM-s, ($N_1$,$N_2$) = (4,4)]
	{\includegraphics[scale=0.4]{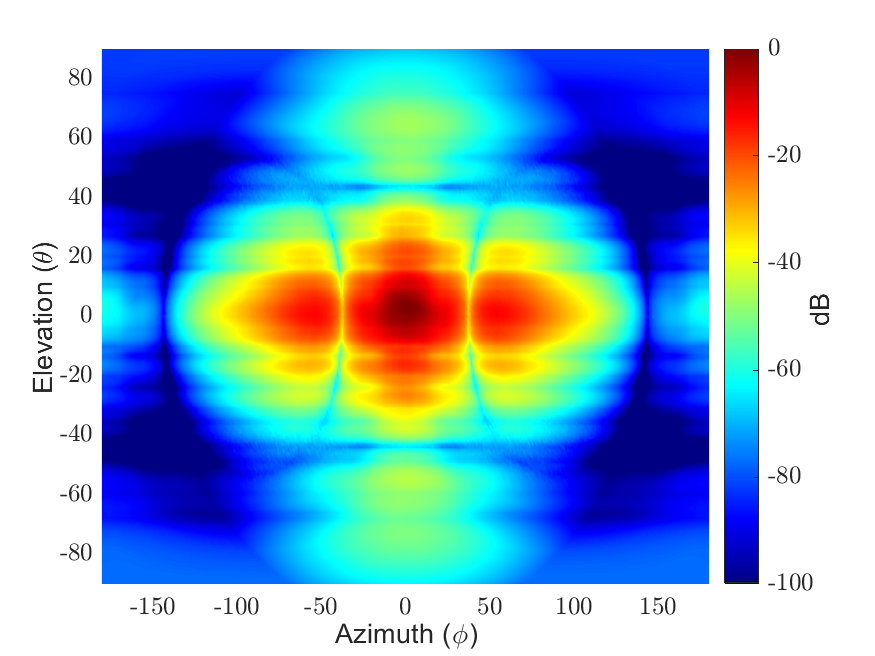}\label{Fig:EIRP-N1-4-N2-4-SSBpmi}}
	\end{subfigure}
    \caption{Average EIRP as a function of azimuth ($\phi$) and elevation ($\theta$) angles, evaluated over all possible PM-s for $N_L$ = 2 under two $(N_1,N_2)$ configurations.}

     \label{Fig:AntPort-vs-EIRP}
\end{figure*}

\begin{algorithm}[t]
\caption{HPBW-based PM Selection}
\label{alg:3dB_selection}
\KwIn{$\mathcal{C}_{\text{panel}}$, $\mathcal{C}_{\text{cb}}$,  $\varepsilon$, $\theta_{i}$, $\phi_{i}$}
\KwOut{$result$}

\SetKwFunction{FMain}{SubsetPM}
\SetKwProg{Fn}{Function}{:}{}
\Fn{\FMain{$\mathcal{C}_{\text{panel}}$, $\mathcal{C}_{\text{cb}}$, $\varepsilon$, $\theta_{i}$,$\phi_{i}$}}{
    $W \leftarrow {\small \texttt{getPMIType1SinglePanelCodebook} (\mathcal{C}_{\text{cb}})}$\; 
    $EIRP \leftarrow \texttt{pattern}(\mathcal{C}_{\text{panel}}, W)$\; 
    $(\theta_{p}, \phi_{p}) \leftarrow \texttt{peak}(EIRP)$\; 
    $\boldsymbol{\theta}_{HPBW} \in \{-\frac{\theta_{HPBW}}{2}+\theta_{p},\frac{\theta_{HBBW}}{2}+\theta_{p}\}$;
    $\boldsymbol{\phi}_{HPBW} \in \{-\frac{\phi_{HPBW}}{2}+\phi_{p},\frac{\phi_{HPBW}}{2}+\phi_{p}\}$\;

    \eIf{$\phi_{i} \notin \boldsymbol{\phi}_{HPBW} ~\&~\theta_{i} \notin \boldsymbol{\theta}_{HPBW}$}{
        $result \leftarrow W$\;
    }{
        \texttt{continue}\;
    }
    \Return{$result$}\;
}
\end{algorithm}

\subsubsection{HPBW-based PM selection}

Algorithm~\ref{alg:3dB_selection} first extracts all Type-I single-panel PMs from the codebook (line 2) and computes their corresponding 3D EIRP patterns based on the antenna panel configuration (line 3). It then determines the peak direction of each beam (line 4) and identifies the azimuth and elevation ranges corresponding to the half-power beamwidth (line 5). Finally, PM-s whose HPBW regions do not overlap with the direction of interest $(\theta_i, \phi_i)$ are selected (lines 6–10), forming the final subset of PM-s.
 \color{black}

\section{Performance Results \& Discussions}
This section presents results in three parts: i) 3D EIRP evaluation for different antenna port configurations, ii) effect of beam nulling via subset-based precoding on average EIRP, and iii) BER performance at the UE for various modulation schemes using full and subset PM sets from the 3GPP predefined codebooks~\cite{gNBH2}. 

\subsection{Impact of Antenna Port Configuration on 3D average EIRP}
All the EIRP results are obtained considering the AAS, SSB and PM described in section~\ref{AASsysmodel}, where we discuss; I) the EIRP generated from all PM-s and II) EIRP composed of all PM-s and a single active SSB with a steering vector at ($\theta_{SSB_i},\phi_{SSB_i}\)). Fig.~\ref{Fig:EIRP-N1-4-N2-4-SSBpmi} illustrates the average EIRP distribution at the gNB as a function of ($\theta,\phi$) angle pair, obtained by averaging over all possible PM sets for $N_L = 2$. The resulting patterns represent the composite radiation characteristics when all precoding configurations are considered. As the number of antenna ports ($P_{CSI-RS}= N_1 \times N_2 \times 2$) increases from  8 to 32, the radiated energy becomes more present along the (\(\theta,\phi\))-plane, indicating enhanced beam directivity in wider directions of both the elevation and azimuth angle. For instance, in case of $(N_1, N_2) = (2, 2)$ with 512 PM-s set, the intense EIRP region is concentrated in the range of $[ -25^\circ, 25^{\circ}]$ for all ($\theta,\phi$) pairs, whereas for $(N_1, N_2) = (4, 4)$ with 2048 PM-s set, it extends up to $[ -50^\circ, 50^{\circ}]$, as shown in Fig.~\ref{Fig:EIRP-N1-8-N2-2} and Fig.~\ref{Fig:EIRP-N1-4-N2-4}, respectively. The wider elevation spread in the latter case indicates increased upward radiation, which may impact the coexistence with radar and airborne incumbents in shared spectrum. On the other hand, lowering the number of ports \((N_1, N_2)\) reduces 3D dimensionality and the PM set size, which limits the ($\theta,\phi$) pair focused directionality, leading to reduced channel exploitation and network capacity. Fig.~\ref{Fig:AntPort-vs-EIRP} shows the impact of SSB configuration on mean EIRP, limiting the PM power distribution towards the steering angle ($\theta_{SSB_i},\phi_{SSB_i}$) of a given SSB.

\subsection{ Impact of The Proposed Beam Nulling Schemes on Average EIRP}

\begin{figure}
	\centering
    \begin{subfigure}[All PM-s]
	    {\includegraphics[scale=.49]{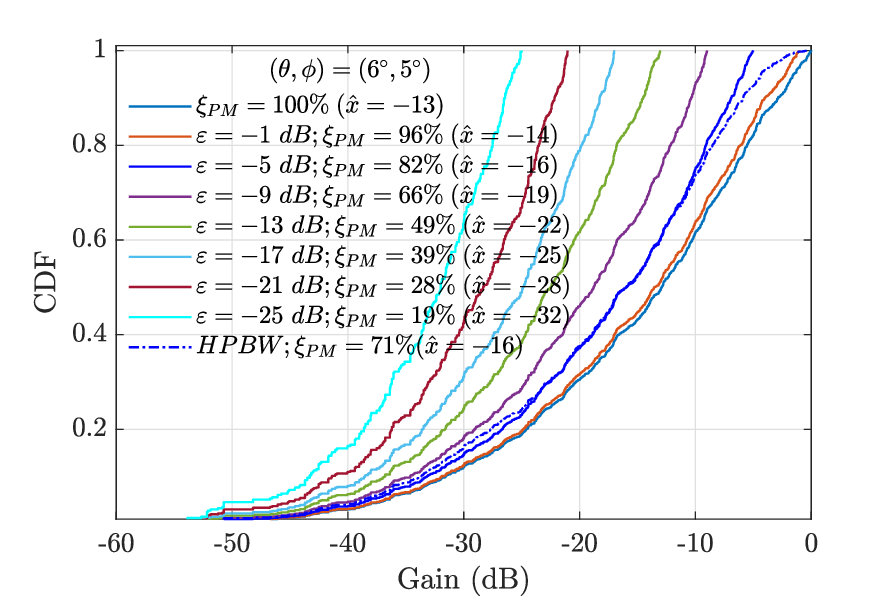}
	    	\label{Fig:cdf_sub1}}
	\end{subfigure}
    \begin{subfigure}[SSB-PM: SSB with conf. \{3, 3, 2\} \& all PM-s ]
	{\includegraphics[scale=0.49]{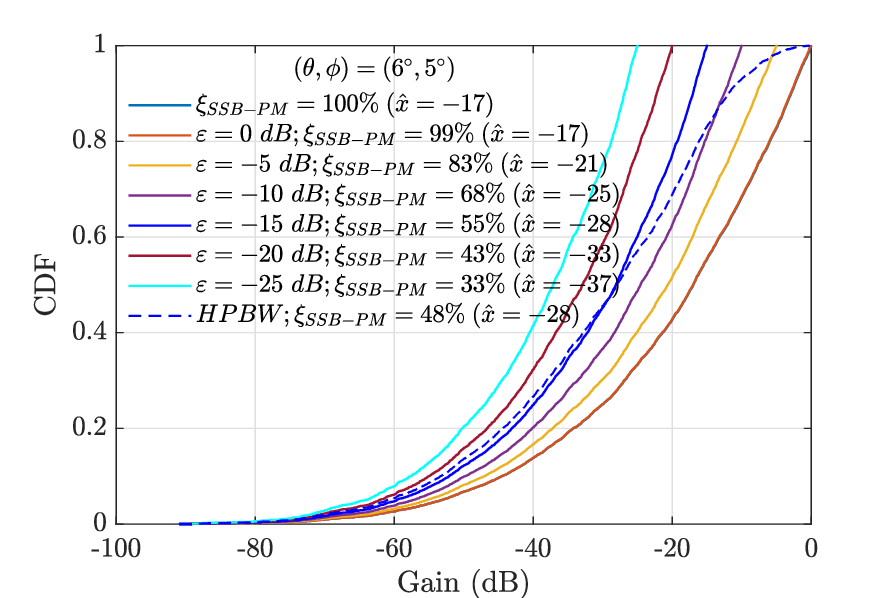}\label{Fig:cdf_sub2}}
	\end{subfigure}
	\caption{CDF of the relative EIRP at ($\theta,\phi) = (6^\circ,5^{\circ}$) considering the proposed beam nulling schemes for ($N_1$, $N_2$) = (4,4) and $N_L$ = 2.}
	\label{Fig:cdf_eirp_44}
\end{figure}

Fig.~\ref{Fig:cdf_eirp_44} shows the cumulative distribution function (CDF) of the normalized EIRP at ($\theta_{i},\phi_{i}) = (6^\circ,5^{\circ}$) for PM set selected using Algorithm~\ref{alg:pm_selection} and Algorithm~\ref{alg:3dB_selection}. 

In Fig.~\ref{Fig:cdf_sub1}, we show the EIRP of all PM-s generated, considering Type I single panel wideband beamforming with $N_1$,$N_2$) = (4,4) and $N_L$ = 2. Up to 3~dB reduction in median EIRP is observed after excluding all PMs whose normalized EIRP value is above $\varepsilon = -5$ dB, at the cost of a reduced subset PM down to 82\% with Algorithm~\ref{alg:pm_selection}. Algorithm~\ref{alg:3dB_selection} discards all PMs with HPBW region ($\boldsymbol{\theta}_{HPBW},\boldsymbol{\phi}_{HPBW}$) encapsulating the direction of interest ($\theta_{i},\phi_{i}$) = ($6^\circ,5^{\circ}$). HPBW-based beam nulling offers the same median EIRP as $\varepsilon$-based nulling with $\varepsilon=-5$ dB at the cost of reduced PM set down to 71\%. $\varepsilon$-based beam nulling limits maximum EIRP at ($\theta_{i},\phi_{i}$) to $\varepsilon$ since it considers both main and side-lobe power. In contrast, HPBW-based nulling allows higher maximum EIRP, considering only power of the main lob. If the required median EIRP is at most -25 dB, we can use $\varepsilon=-17$ dB with reduced PM set down to 38.7\%.

In Fig.~\ref{Fig:cdf_sub2}, we observe a 4 dB lower median EIRP under the combination of PM-s with one of the SSB-s, SSB-PMI scenario, compared to using only PM-s. Moreover, the median EIRP lowers to -28 dB with HPBW-base selection scheme at the cost of a reduced PM set down to 48\%. $\varepsilon$-based scheme with $\varepsilon=-15$ offers same median EIRP with HPBW scheme with 7\% larger selection set. 

\begin{figure}
	\centering	\includegraphics[scale=0.5]{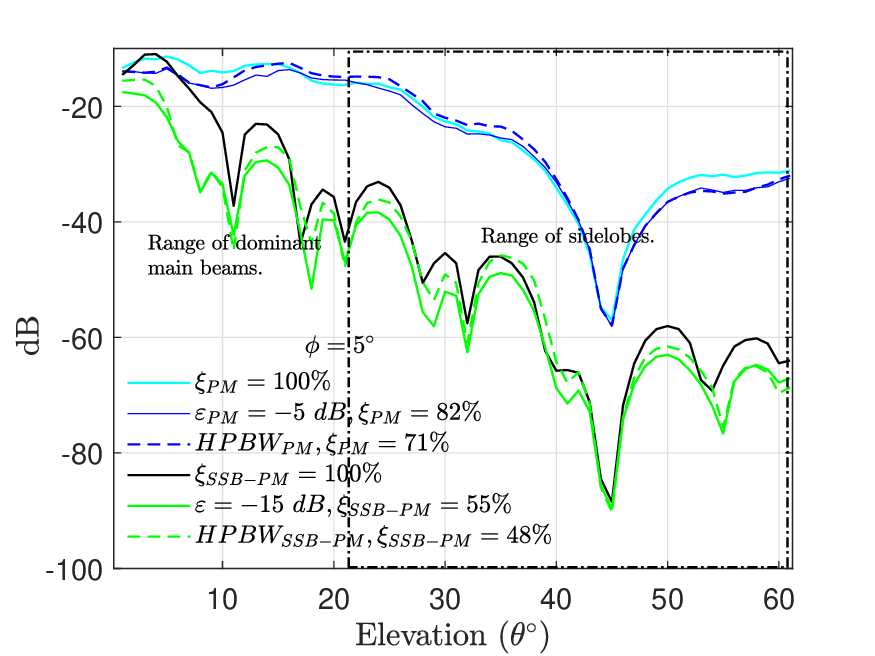}
	\caption{Median level of normalized EIRP considering all PM-s and the proposed beam nulling schemes for ($N_1$,$N_2$) = (4,4) and $N_L$ = 2.}
	\label{Fig:Median_Diff_EIRP_subset}
\end{figure}

Fig.~\ref{Fig:Median_Diff_EIRP_subset} shows the median EIRP of all PM and SSB-PM scenario versus the proposed beam nulling schemes at azimuth cut (\(\phi=5^{\circ}\)). We observe up to 15 dB lower median EIRP along elevation for both main and side-lobe region via the proposed beam nulling schemes. The median EIRP of SSB-PM is lower and more controllable than only PM scenario due to SSB masking.

\subsection{Beam Nulling and BER Performance Trade-Offs}

In this section, the BER performance has been evaluated using an end-to-end 5G~NR downlink single-user MIMO link-level simulator described in subsection \ref{subsec:system_model}, considering; i) SVD-based precoding derived directly from the channel matrix, ii) standard 3GPP Type I single-panel codebook selection PM, and iii) proposed beam nulling-based PM selection schemes. The key configurable system parameters for the Monte-Carlo simulator are summarized in Table~\ref{tab:linkSimParams}. BER results are shown only for SNR values where the packet error rate exceeds 1\%, ensuring statistically significant error counts for reliable estimation. Under perfect channel, precoding and equalization use perfect CSI; under imperfect channel, they rely on channel estimation.

\begin{table}[t]
\caption{Link-Level Simulation Parameters}
\renewcommand{\arraystretch}{1.05}
\centering
\begin{tabular}{|p{0.48\columnwidth}|p{0.42\columnwidth}|}
\hline
\multicolumn{2}{|c|}{\textbf{Transmission Parameters}} \\ \hline
Modulation schemes & 16-, 64-, 256-QAM \\ \hline
$N_L$, Tx/Rx antennas & 2, 32/4 \\ \hline
Coding rate, iterations & 490/1024, 10{,}000 \\ \hline
\hline
\multicolumn{2}{|c|}{\textbf{Carrier Configuration}} \\ \hline
Subcarrier spacing, RBs, BW & 30~kHz, 52, 20~MHz \\ \hline
\hline
\multicolumn{2}{|c|}{\textbf{Channel Model}} \\ \hline
Type, delay spread, Doppler & TDL-C (Urban Macro NLOS), 300~ns, 0~Hz \\ \hline
\hline
\multicolumn{2}{|c|}{\textbf{CSI-RS Configuration}} \\ \hline
Ports, row, periodicity & 32, 16, 4 slots \\ \hline
\hline
\multicolumn{2}{|c|}{\textbf{PDSCH Configuration}} \\ \hline
DMRS type/len./pos. & Type~1 / 2~sym. / +1 pos. \\ \hline
PRB allocation & Full band (0--51) \\ \hline
\hline
\multicolumn{2}{|c|}{\textbf{Codebook Configuration}} \\ \hline
Type, panel ($N_1, N_2$) & Type~I Single Panel, (4,4) \\ \hline
PMI/CQI mode & Wideband / Wideband \\ \hline
Precoding schemes & SVD, PMI, $\varepsilon=-15$ dB threshold, HPBW\\ \hline
\hline
\multicolumn{2}{|c|}{\textbf{Decoder Parameters}} \\ \hline
LDPC algorithm, max iters & Norm. min-sum, 8 \\ \hline
\end{tabular}
\label{tab:linkSimParams}
\end{table}

\begin{figure}
	\centering	\includegraphics[scale=0.5]{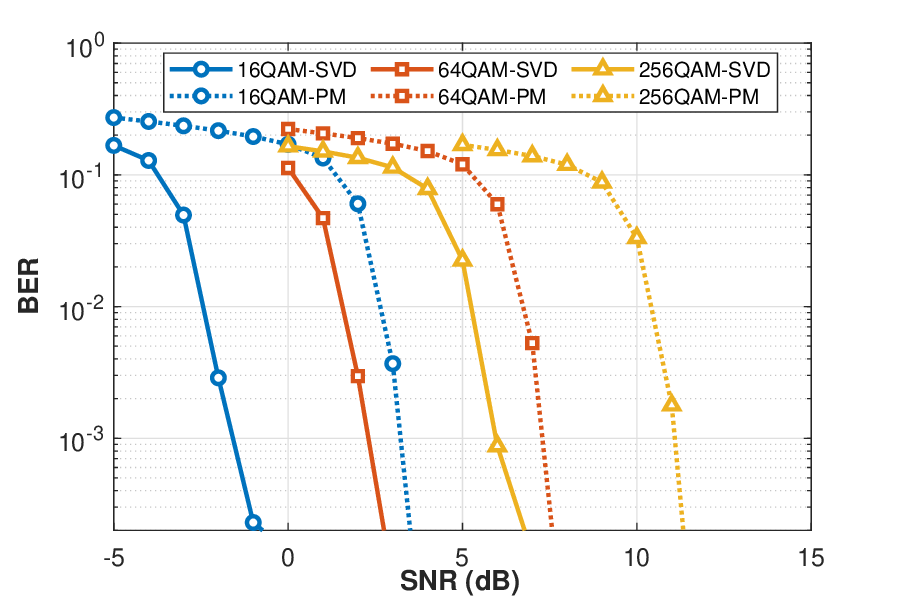}
	\caption{BER performance of PM vs SVD-based precoding considering ($N1$,$N2$) = (4,4) and $N_L$ = 2. }
	\label{Fig:ber_n1n2_44_l2_PMI_vs_SVD}
\end{figure}

Fig.~\ref{Fig:ber_n1n2_44_l2_PMI_vs_SVD} illustrates the BER performance of the UE using the standard PM and SVD-based precoding under various modulation schemes, while considering a perfect channel estimation. As expected, the 3GPP-specific precoding yields poorer performance since it is not optimized for the UE’s channel. In this case, the UE-specific precoding is performed via a fixed PM-s predefined in the 3GPP standards. For example, at a BER of \(10^{-3}\), SVD-based precoding outperforms standard 3GPP PM-based precoding with 16-QAM by approximately 5 dB in SNR. Moreover, the performance gap between the two methods widens with higher modulation orders.

\begin{figure}
	\centering
	\includegraphics[scale=0.48]
{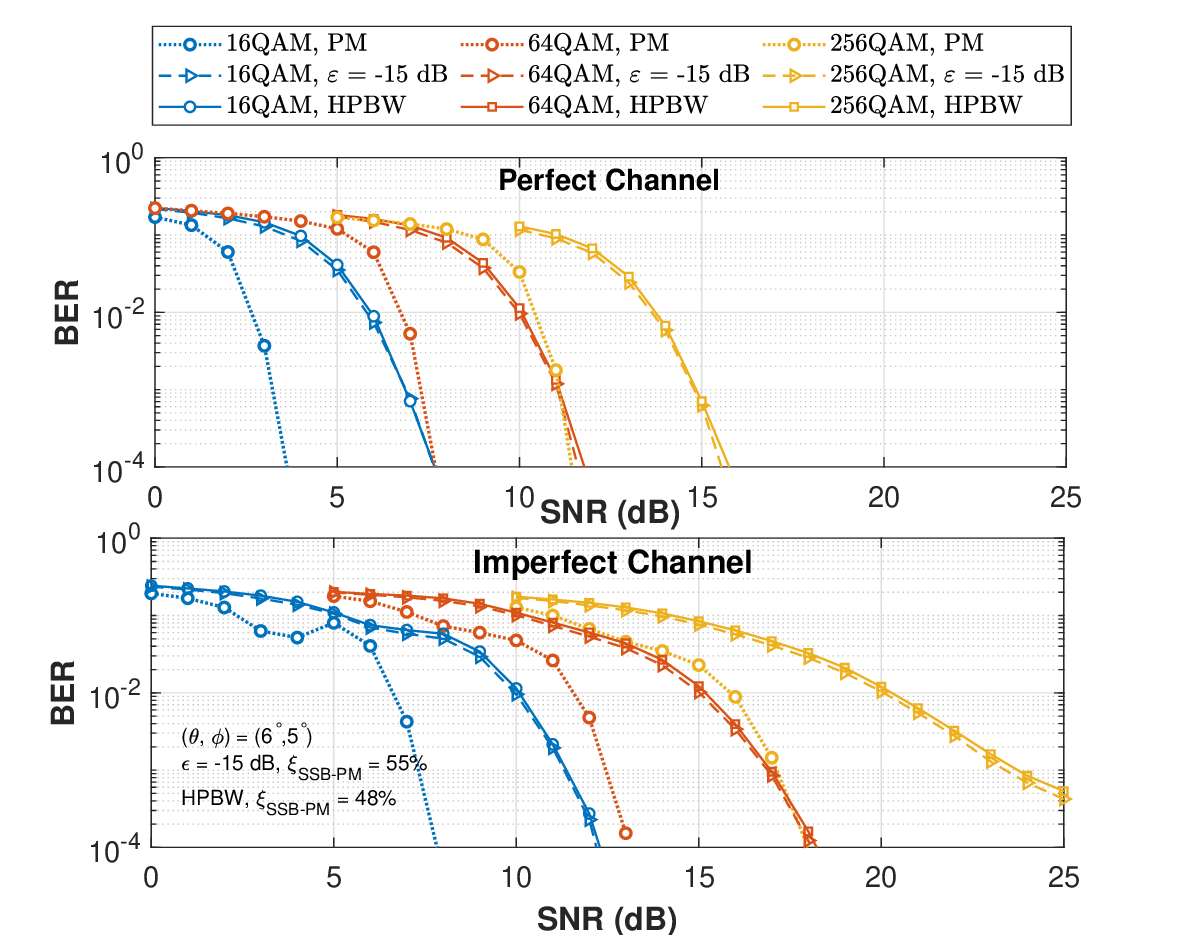}
	\caption{BER performance for ($N1$,$N2$) = (4,4) on PDSCH and $N_L$ = 2.}
	\label{Fig:ber_n1n2_44_l2_pmiSvd}
\end{figure}

Fig.~\ref{Fig:ber_n1n2_44_l2_pmiSvd} compares the BER performance of the SSB-PM and the proposed beam nulling-based precoding schemes ($\varepsilon$ threshold and HPBW). Due to the reduced set of available PM-s by approximately 50\%, discussed for direction ($\theta_{i},\phi_{i}) = (6^\circ,5^{\circ}$) with reduced median EIRP by 11 dB, the proposed beam nulling approaches with 16QAM exhibit about 3.5 dB SNR degradation performance at \(10^{-4}\) BER. Under the same BER, SNR degradation becomes 3.5 dB and 4.5 dB for 64QAM and 256QAM, respectively. Only a marginal BER improvement is observed, even though the threshold-based beam nulling precoder retains 7\% more available PMs than the HPBW-based approach. This discussion highlights the trade-off between lowering the gNB’s transmitted EIRP towards the direction of interest at ($\theta_{i},\phi_{i}$) and maintaining the UE’s link performance. 

\section{Conclusion}
This work investigates the coexistence of a 3GPP-compliant 5G system with airborne incumbents by analyzing the gNB’s 3D normalized EIRP distribution and the UE’s BER performance across four precoding techniques: optimal SVD-based precoding, 3GPP-defined precoding, and two proposed beam nulling-based precoding methods. The results demonstrate that the proposed beam-nulling schemes achieve 11 dB reduction in radiation toward the target direction, costing approximately 3.5–4.5 dB of SNR loss in UE link performance across multiple modulation schemes when targeting a BER of $10^{-4}$. The 3D EIRP distribution is heavily dependent on the AAS architecture, antenna port configuration, and underlying system parameters employed, and further studies are being performed to comprehensively characterize these dependencies. Our future work will also focus on incorporating insights from real-world 5G measurement campaigns into the simulation framework to enhance accuracy and extend the analysis to additional precoding schemes introduced in 3GPP Release 18.


\bibliographystyle{IEEEtran}
\bibliography{ReferenceList}

@techreport{ericsson_aas_5g_2023,
  author       = {{Ericsson}},
  title        = {{Advanced antenna Systems for 5G networks}},
  institution  = {Ericsson AB},
  year         = {2023},
  note         = {White Paper, Available: \url{https://www.ericsson.com/en/reports-and-papers/white-papers/advanced-antenna-systems-for-5g-networks}}
}

@techreport{fcc_upper_cband_consensus_2025,
  author       = {{Wireless and Aviation Industry Representatives}},
  title        = {{Upper C-band technical consensus: Wireless \& aviation joint industry presentation}},
  institution  = {Federal Communications Commission (FCC)},
  year         = {2025},
  month        = sep,
  note         = {Meeting held on September 30, 2025. Available: \url{https://www.fcc.gov/ecfs/document/10022930930436/1}}
}

@standard{3gpp_tr_38901_2022,
  title        = {{Study on channel model for frequencies from 0.5 to 100 GHz}},
  organization = {3rd Generation Partnership Project (3GPP)},
  institution  = {ETSI},
  number       = {TR 38.901 V17.0.0},
  year         = {2022},
  month        = dec,
  note         = {Available: \url{https://www.3gpp.org/DynaReport/38901.htm}}
}

@online{idst_aas_breakthroughs_2022,
  author       = {{International Defence, Security \& Technology (IDST)}},
  title        = {{Antennas and advanced antenna systems (AAS) breakthroughs to accelerate large-scale deployments of 5G networks}},
  year         = {2022},
  note         = {Available: \url{https://idstch.com/technology/electronics/advanced-antenna-breakthroughs-to-accelerate-deployments-of-5g-networks/}}
}

@ARTICLE{9681631,
  author={Lin, Xingqin and Cioni, Stefano and Charbit, Gilles and Chuberre, Nicolas and Hellsten, Sven and Boutillon, Jean-Francois},
  journal={IEEE Communications Magazine}, 
  title={{On the path to 6G: Embracing the next wave of low earth orbit satellite access}}, 
  year={2021},
  volume={59},
  number={12},
  pages={36-42},
  doi={10.1109/MCOM.001.2100298}}

@techreport{5gamericas_midband_spectrum_2021,
  author       = {5G Americas},
  title        = {{Mid-Band spectrum and the co-Existence with radio altimeters}},
  institution  = {5G Americas},
  year         = {2021},
  month        = {Jul.},
  note         = {Available: \url{https://5gamericas.org/wp-content/uploads/2021/07/Mid-Band-Spectrum-and-the-Co-Existence-with-Radio-Altimeters.pdf}}
}

@ARTICLE{10459211,
  author={Kang, Seongjoon and Mezzavilla, Marco and Rangan, Sundeep and Madanayake, Arjuna and Venkatakrishnan, Satheesh Bojja and Hellbourg, Grégory and Ghosh, Monisha and Rahmani, Hamed and Dhananjay, Aditya},
  journal={IEEE Open Journal of the Communications Society}, 
  title={{Cellular wireless networks in the upper mid-band}}, 
  year={2024},
  volume={5},
  number={},
  pages={2058-2075},
  doi={10.1109/OJCOMS.2024.3373368}}

@standard{3gpp_38101_1,
  title        = {{NR; User Equipment (UE) radio transmission and reception; Part 1: Range 1 Standalone}},
  organization = {3rd Generation Partnership Project (3GPP)},
  institution  = {ETSI},
  number       = {TS 38.101-1 V17.15.0},
  year         = {2024},
  month        = {Nov.},
  note         = {{Available: \url{https://www.3gpp.org/DynaReport/38101-1.htm}}}
}

@article{bogale2016massive,
  title={{Massive MIMO and mmWave for 5G wireless HetNet: Potential benefits and challenges}},
  author={Bogale, Tadilo Endeshaw and Le, Long Bao},
  journal={IEEE Vehicular Technology Magazine},
  volume={11},
  number={1},
  pages={64-75},
  year={2016},
  publisher={IEEE}
}

@article{ross2019annual,
  title={{Annual report on the status of spectrum repurposing}},
  author={Ross, Wilbur L and Kinkoph, Secretary Douglas W},
  journal={Proceedings of the 2019, US Department of Commerce},
  year={2019}
}

@inproceedings{farshchian2023modeling,
  title={Modeling and Impact of Cellular Uplink Aggregate Interference on Radar Performance},
  author={Farshchian, Masoud and Zebrowitz, Harris and Baker, Amy and Howard, Fredrick},
  booktitle={2023 IEEE Radar Conference (RadarConf23)},
  pages={1-6},
  year={2023},
  organization={IEEE}
}

@article{gNBH2,
  title={{5G NR: Physical layer procedures for data (Release 17)}},
  author={3GPP},
  journal={document TS 38.214, V17.0.0},
  year={2022},
 month = {Mar.}
}

@article{jo2011transmit,
  title={{Transmit-nulling SDMA for coexistence with fixed wireless service}},
  author={Jo, Han-Shin and Mun, Cheol},
  journal={Journal of the Korean Institute of Electromagnetic and Science},
  volume={11},
  number={1},
  pages={34-41},
  year={2011},
  publisher={The Korean Institute of Electromagnetic Engineering and Science}
}

@inproceedings{kang2024terrestrial,
  title={{Terrestrial-satellite spectrum sharing in the upper mid-band with interference nulling}},
  author={Kang, Seongjoon and Geraci, Giovanni and Mezzavilla, Marco and Rangan, Sundeep},
  booktitle={ICC 2024-IEEE International Conference on Communications},
  pages={5057-5062},
  year={2024},
  organization={IEEE}
}

@article{jia2025joint,
  title={{Joint Detection, Channel Estimation and Interference Nulling for Terrestrial-Satellite Downlink Co-Existence in the Upper Mid-Band}},
  author={Jia, Shizhen and Ying, Mingjun and Mezzavilla, Marco and Calin, Doru and Rappaport, Theodore S and Rangan, Sundeep},
  journal={arXiv preprint arXiv:2510.08824},
  year={2025}
}

@book{asplund2020advanced,
  title={{Advanced antenna systems for 5G network deployments: bridging the gap between theory and practice}},
  author={Asplund, Henrik and Karlsson, Jonas and Kronestedt, Fredric and Larsson, Erik and Astely, David and von Butovitsch, Peter and Chapman, Thomas and Frenne, Mattias and Ghasemzadeh, Farshid and Hagstr{\"o}m, M{\aa}ns and others},
  year={2020},
  publisher={Academic Press}
}


\end{document}